# Teacher Learning of Technology-Enhanced Formative Assessment

Ian D. Beatty, Allan Feldman, William J. Leonard, William J. Gerace,
Karen St. Cyr, Hyunju Lee, Robby Harris

*Scientific Reasoning Research Institute*
*University of Massachusetts Amherst*

*Technology-Enhanced Formative Assessment* (TEFA) is a pedagogy for teaching with classroom response technology. *Teacher Learning of TEFA* is a five-year research project studying teacher change, in the context of an intensive professional development program designed to help science and mathematics teachers learn TEFA. First, we provide an overview of the project's participating teachers, its intervention (consisting of the technology, the pedagogy, and the professional development program), and its research design. Then, we present narratives describing the unfolding change process experienced by four teachers. Afterward, we present some preliminary findings of the research, describe a "model for the co-evolution of teacher and pedagogy" that we are developing, and identify general implications for professional development.

This paper is a slightly elaborated version of the "script" for our presentation at the 2008 conference of the *National Association for Research in Science Teaching* in Baltimore, Maryland, USA. It was presented in Strand 4 ("Science Teaching—Secondary School") from 2:00 to 3:30 on Tuesday, April 1. We have deliberately retained an informal, conversational manner here, and tried to mimic a little of the accompanying slides' style.

The project and its presentation have been funded by the US National Science Foundation through its Teacher Professional Continuum program, grant number TPC-0456124.

## part 0: introduction

We are interested in **teacher change**: in promoting it, supporting it, and understanding it. That means we're interested in the **process** of teacher change — how teachers change their views, their capacities, their practices, and their "way of being a teacher" (Blum, 1999; Davis, Feldman, Irwin, Pedevillano, Capobianco & Weiss, 2003).

Abundant evidence documents the **gap** between the science education research community's general knowledge of what makes for effective instruction and what happens every day in many classrooms (Bransford, Brown & Cocking, 1999). There is also abundant evidence that most teachers really **want** to do their jobs as well as possible, but that pedagogical change is hard, and most interventions fail, or at least succeed only weakly compared to the aspirations of their designers (Kennedy, 2006).

We don't think enough is known about **how** teacher change occurs, **why** it is difficult, or **how** we can better facilitate it.

*Teacher Learning of Technology-Enhanced Formative Assessment* (TLT) is a five-year research project focused on studying teacher learning and pedagogical change. Our goals today are to tell you a little bit about the project, to give you a glimpse through our window into teacher change, and to share some of our preliminary findings.

The core of this presentation, in part 2, consists of four **stories**: narratives of teacher change for four of our participating teachers. But first, in part 1, we'll describe the **context** of those stories: our project participants, our intervention, and our research design. At the end, in part 3, we'll describe some general **findings** that cut across these four cases and others, present a model of teacher change that we've been developing, and identify some implications for professional development.

# part 1: context

To study teacher change, we need three ingredients: **teachers**, an **intervention** that induces change, and a **research design** for observing that change.

## a. the teachers

We are working with approximately 39 teachers from three school systems. We say "approximately" because there is some attrition, and the number depends on exactly **when** you count. The total number about whom we have at least a little data is larger; the number currently active is smaller.

We have four "sites" in three school districts, all in western Massachusetts. We've staggered the start dates to distribute the load on our professional development and research staff, and to provide us the opportunity for mid-course corrections.

Our first site is a combined middle-and-high school in a **rural multi-town district**. For a while it was flirting with a "high needs" designation, but seems to have escaped that. We began our first intervention year there — in August 2006 — with ten participating teachers, six from the high school and four from the middle school. Four taught science (two at each level), and the remainder taught math. After the first intervention year ended, four teachers left the project for various reasons, although two indicated that they wanted to continue practicing the pedagogy, but without participating in professional development or data collection activities.

Our second site, begun one year later, is a high school in a **diverse college town**. These teachers were so enthusiastic that they basically pestered us into letting them join; we accommodated them by switching from a two-site to a three-site design. All seven teachers of the science department are participating, as well as one math teacher.

Our third and fourth sites have just recently been organized, and the intervention will begin in August 2008. The third site is one of the largest high schools in western Massachusetts, in a **sprawling suburban area** adjacent to a large city. We have ten science teachers and one math teacher participating, as well as one science teacher from the nearby vocational school.

As we were negotiating with that school, the district's middle school teachers asked to join. The opportunity to learn more about how our pedagogy and professional development work at the middle school level, and to impact an entire district's secondary schools at once, was too good to pass up. So, we combined the interested teachers from the **two middle schools** into a fourth "site" that will run in parallel with the third site. This fourth site has eight teachers, all in science.

(This causes our total number of participants to significantly exceed the number we planned on. Due to unforeseen economies, our budget can handle this, but it will place a severe strain on our person-power, and we will be forced to become more efficient. This is not necessarily bad, since we want to look down the road towards scalability, anyway.)

So, you can see that we have quite a mix of math and science, middle and high school, although we're slanted towards high school science since that was our original focus and mandate.

We also have a cross-section of socioeconomic types. It's not quite as broad as we wanted: we tried hard to find an urban school with a large minority population that we could partner with, and we came close, but the ones we negotiated with couldn't pull together the basic technical infrastructure that the project required. Which, of course, simply highlights the difficulties facing that kind of school district.

### b. the intervention

The second ingredient of our study is the intervention. Our intervention consists of a **hook**, a **pedagogy**, and a **professional development** program.

#### *the hook*

The hook is the bit that catches teachers' attention and captures their initial interest. It's not a shallow gimmick, but rather something distinctive and characteristic that they can immediately see potential in. Our hook is the use of ***classroom response system*** (CRS) technology, known informally as "clickers" (Beatty, 2004; Dufresne, Gerace, Leonard, Mestre & Wenk, 1996; Fies & Marshall, 2006).

In essence, a CRS is nothing more than a set of simple transmitters that students use to send in their answers to some question; a receiver and software that runs on the teacher's computer, collecting and instantly aggregating answers from the whole class; and a way to display the distribution of answer choices to the teacher and the students, typically as a histogram on a data projector or wall-mounted monitor. In our experience, CRSs are becoming increasingly common in universities (Banks, 2006), but are still rare in K-12 schools.

A CRS is just a tool, and a teacher can use that tool for many, many different purposes. This raises the question of what a teacher **should** use one for. And that is where our pedagogy comes in.

### *the pedagogy*

The "pedagogy" part of our intervention is something we've been developing since 1993. It began in the context of university physics, expanding to other subjects and into secondary schools. We call it **Technology-Enhanced Formative Assessment** — "TEFA" for short. It's quite rich and rather carefully designed, and we could easily fill a ninety-minute session talking about just that. So, we'll barely scratch the surface here. A more detailed description of some of its components and aspects is available in previous writings (Dufresne et al., 1996; Beatty, Leonard, Gerace & Dufresne, 2006), and we are currently preparing a definitive exposition and defense for journal publication.

At the heart of TEFA, defining and directing it, are **four core principles**: *question-driven instruction*, *dialogical discourse*, *formative assessment*, and *meta-level communication*.

1. Motivate and focus student learning with **question-driven instruction** (QDI). This means posing tough, rich, meaty, often messy questions to students in order to contextualize and motivate subsequent learning, and often in order to catalyze or precipitate learning. It is grounded in the conceptual change tradition (Scott, Asoki & Leach, 2007). It is motivated by an understanding that students perceive, process, and store information differently in response to a need, and that they "get" ideas by wrestling with the application of those ideas (Bransford et al., 1999, p. 139).

2. Develop students' understanding and scientific fluency with **dialogical discourse** (DD). This means engaging students in discussion in which multiple ideas and ways of thinking are explored and contrasted, and in which students articulate and explore their own thinking. It is grounded in the sociocultural tradition (Carlsen, 2007; Mortimer & Scott, 2003). It is motivated by an understanding that learning science largely means learning the *social language* of science (Bakhtin, summarized in Wertsch, 1991, pp. 93-118), and one must practice speaking a language to develop fluency. It is also motivated by an understanding that the tools students use for internal cogitation are appropriated from social interactions (Vygotsky, 1987).

3. Optimize teaching and students' learning with **formative assessment** (FA) In this context, this means making students' knowledge and thinking visible in order to adjust and optimize subsequent learning and teaching. It is motivated by an understanding that effective instruction requires detailed and current information about the specific students being taught, and that effective learning requires accurate self-knowledge (Wiliam, 2007). According to a seminal literature review by Paul Black and Dylan Wiliam (1998), educational "innovations" involving formative assessment produce learning gains that are among the largest ever found for educational innovations.

4. Help students cooperate in the learning process and develop metacognitive skills with **meta-level communication** (MLC). This means communicating about communication, about cognition, about learning, and about the purposes of instructional experiences. It is grounded in literature on student motivation and self-regulation (Koballa & Glynn, 2007; Wilson, 2006). It is motivated by an understanding that learning works better when students frame their participation appropriately and understand what they are supposed to be paying attention to (Hammer, 1996; Hammer & Elby, 2003).

These four principles are *not* independent and arbitrary beliefs. Rather, they interlock and reinforce each other in a highly synergistic way. This can be seen in the way they come together in the *question cycle*, the canonical or prototypical way that TEFA is enacted in the classroom (figure 1).

The teacher begins a cycle by presenting a **question** or problem to the class. Students **think** about it, ether individually or talking in small groups, and and decide upon their answers. They then enter their **responses** into their classroom response system transmitters. The software aggregates the answers and constructs a **histogram** indicating how many students have picked each answer, which the teacher shares with the class.

Then, the teacher moderates a whole-class **discussion** about the question. This typically begins by asking for volunteers to explain the reasoning behind their particular choice. "I see someone chose 'not enough information.' Could you tell me what else you would like to know in order to solve it?" "Several people picked answer three. Who can give me an argument why that's a good choice?" "Hm, okay. Did anyone have a *different* reason for choosing that same answer?"

The teacher's initial goal is to draw out the range and diversity of thinking behind different answers, in as much depth and rigor as possible, and then to evolve the discussion of these into examination and comparison of the ideas, exploration of related topics, and development of understanding or insight. Finally, the teacher ends the cycle by providing some kind of summary, micro-lecture, or other **closure**, and then transitioning to another question or a different activity (or perhaps ending class).

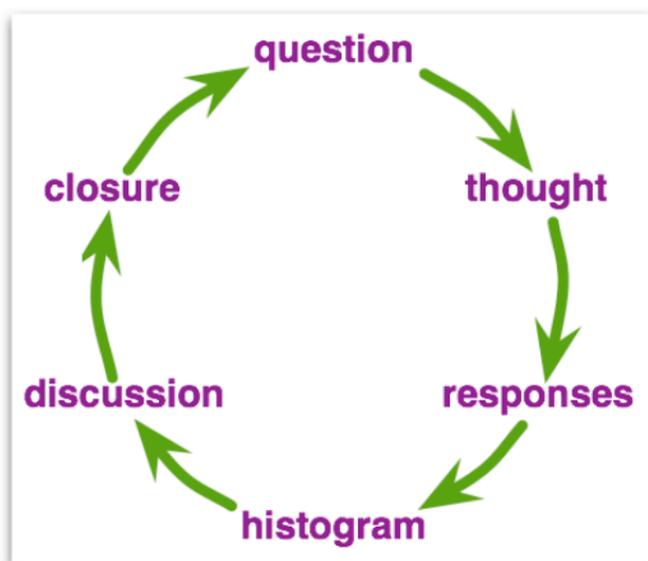

*Figure 1: The TEFA "question cycle".*

We often use a sequence of related questions to develop students' understanding of some set of ideas, homing in on the desired learning through multiple iterations of the question cycle.

It is important to understand that TEFA, enacted in the question cycle, is not meant as a "drop-in" for occasionally quizzing students about what they are supposed to have learned already. It is intended to contextualize, motivate, and actually precipitate student learning, as well as to keep the teacher well informed about students' understanding and thinking. We do not mean that a K-12 teacher should use TEFA for all of their class time every day, but we do advocate it as a regular **engine of learning**, something used frequently and predictably to drive instruction, with other activities surrounding it and relating to it.

### *the professional development program*

Our intervention includes a professional development program designed to help our participating teachers learn to use the CRS technology and implement the TEFA pedagogy. In a sense, the intervention *is* the PD program, although it is inseparable from the technology and pedagogy.

TEFA isn't lightweight, and neither is our PD. We've tried to design in **known best practices** from the in-service teacher PD literature (Supovitz & Turner, 2000; Ball & Cohen, 1999; NEIRTEC, 2004). As we gain experience implementing it, we make improvements for each subsequent site.

Because we're working with several teachers at a time from one school (or from nearby schools with a close administrative relationship), we can provide all PD **on-site,** at the teachers' own schools. This is efficient for the teachers, and we expect it to reduce absenteeism at PD meetings.

Because we work with a critical mass of teachers from a school, and especially from its science department, we can foster a **community of peers** who learn together, act as a problem-solving resource for each other, and sometimes collaborate on question design (curriculum).

The PD program is **intensive**. We begin with a four-day workshop in August, where we paint the "big picture" of TEFA, provide some hands-on time using the technology and creating questions to use with their classes, and generally help them get ready to try the approach out when the school year starts. During the first academic year, we meet once a week, after school, for an hour and a half.

For the second and third academic years of the intervention (second year only at sites 3 and 4), we meet after school every two to three weeks to provide **sustained**, ongoing support. Our experience from previous projects is that it typically takes about three years for most teachers to really internalize TEFA (Feldman & Capobianco, 2008).

The three-year program starts intensively and with a strong facilitator-driven agenda. During years two and three, our focus shifts to **scaffolding**, switching to a teacher-driven mode in order to develop sustainable habits of ongoing professional growth. This is done by involving the teachers in a form of collaborative action research known as *enhanced normal practice* (Feldman, 1996; Feldman & Capobianco, 2000; Feldman & Minstrell, 2000).

We try to be **self-consistent**, using the TEFA methodology to teach TEFA. We do this because we believe it works, and also so that teachers get to see the pedagogy modeled and experience it as students.

The PD remains **grounded** in teachers' real-world classroom experiences. At every meeting, participants discuss their current problems, insights, observations, and ideas. This helps the teachers connect the content of PD with their practice.

It also helps PD staff keep the program **agile**. Because every teacher's outlook, context, and learning trajectory are different, and because each school has its own characteristics, we keep PD highly responsive to our participants' evolving needs. We solicit feedback through multiple channels and make frequent adjustments.

In previous work (Feldman & Capobianco, 2008), we have identified **five general skill areas** that a teacher must develop to master TEFA and be comfortable with it.[1] The general scope of the PD program follows this framework.

The most obvious set of skills to master are related to the **technology**: those required to reliably operate the CRS hardware and software, manage the logistics of the classroom, and deal gracefully with any glitches that arise.

The second skill area addresses crafting **questions** for TEFA question cycles that reliably produce good classroom interaction and achieve learning goals (Beatty, Gerace, Leonard & Dufresne, 2006) — and doing it without consuming more preparation time than the teacher can afford. Teachers quickly discover that typical quiz, homework, and end-of-chapter questions rarely lead to satisfactory TEFA experiences.

The third skill area involves orchestrating the **discourse** of the classroom: encouraging students' participation, eliciting thoughtful and extended contributions, drawing out multiple points of view, and gently steering discussion in order to achieve instructional goals.

The fourth skill area centers on understanding and reacting to **students**: probing their knowledge and thinking through QDI questions and verbal interaction, interpreting their responses and comments, building and refining mental models of the students, and making sound and "agile" adjustments — often in real-time — to the information provided by formative assessment. Formative assessment is only "formative" if the information is actually used to adjust teaching or learning (Black, Harrison, Lee, Marshall, & Wiliam, 2002).

---

[1] Most of our previously published work identifies *four* skill areas; we've since expanded that to five, adding one we had neglected.

The fifth skill area includes everything necessary to integrate TEFA practice with everything else in a teacher's **context**: their curriculum and syllabus, other instructional activities, students' needs and capacities, impending standardized exams, scheduled and unscheduled disruptions, and everything else that fills a teacher's life. The "skills" learned to deal with these are idiosyncratic to each school and teacher's particular circumstances, which means they're hard for us to teach. Also, these kinds of problems don't tend to stay "solved", since circumstances change.

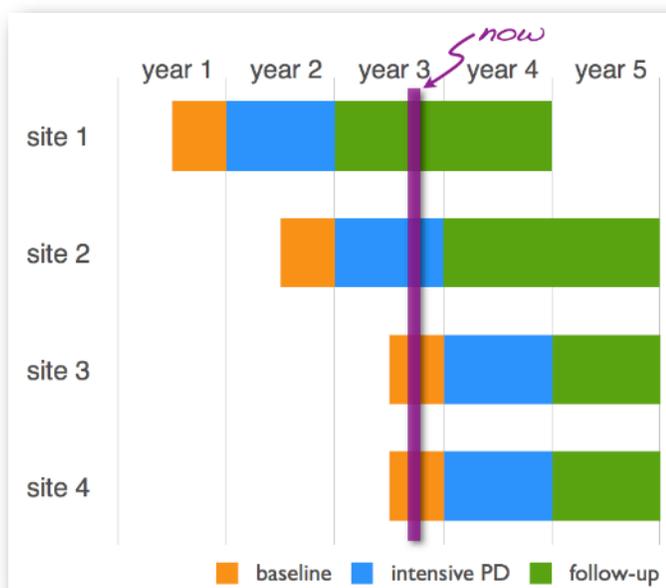

*Figure 2: Project timeline.*

### c. the research design

The third ingredient of our project is the research design: the part that examines how teachers are (hopefully) changing in response to the intervention.

We are using a **longitudinal**, **staggered site**, **delayed intervention** design (figure 2). For each of our four sites, we start with one semester for collecting baseline data, followed by three years of intervention and collection of longitudinal data. Each cohort begins one year after the previous one, both to distribute our workload and to give us a chance to improve our intervention, instrumentation, and methods as we go. (The final two cohorts, consisting of the high school and middle schools from the same district, are running simultaneously.)

As this is written, we are approximately 1.5 semesters into the action research phase at site 1, 1.5 semesters into the intensive PD phase at site 2, and part-way through baseline data collection at sites 3 and 4. In terms of analysis, we've recently collected enough longitudinal data on the teachers at our first site to begin constructing and pondering reasonable case study profiles and starting cross-case analysis. (We've been making observations and kicking around ideas all the time, of course, but now we can get more formal and serious about it.)

#### data sources
We are collecting data through multiple instruments and methods.

**PD meetings:** We audio-record and transcribe all professional development meetings, and take field notes as well.

**journals:** As part of PD, teachers are encouraged to write journal entries on their TEFA thinking and learning. The frequency and quality of these varies drastically from teacher to teacher, however: journaling is sufficiently time-intensive that we don't feel we can insist on it.

**implementation log:** Teachers complete a simple, 2-page paper log form every day, indicating whether they used TEFA, how many questions they posed, how satisfied they were with the outcome, what fraction of the class participated in discussions, what fraction of class time was spent on TEFA, and whether any technical problems arose.

**monthly survey:** Once a month, teachers go online to complete a fairly extensive web-based survey about their experiences with TEFA over the preceding month. It combines multiple choice scales and short essay questions, and some of the questions require significant reflection.

**class video:** Twice per semester, we videotape each teacher teaching an entire class, scheduling — if possible — a day on which they use TEFA. The protocol includes short pre- and post-class interviews to provide us with context and insight.

**student survey:** Once per semester, we administer a Likert scale, optically scanned survey to every student in every class of each participating teacher. The survey has been designed to solicit students' perceptions of the class environment and the teacher's expectations, so we can see whether use of TEFA impacts the learning environment in a way students notice.

**pedagogical perspectives survey & interview:** Once a year, teachers take a Likert scale survey on their general pedagogical views, and an accompanying intensive interview (two parts of approximately 45 minutes each) on perspectives more directly related to TEFA.

**biography:** During the baseline phase, we elicit aspects of each teacher's professional preparation and background through a web-based survey.

**miscellaneous:** Finally, we do some more spontaneous interviews and focus-group discussions on an as-needed, as-possible basis, in order to gain additional insight into teachers' experiences as they encounter and learn TEFA.

### *analysis*

Analysis of all this data is what one might call "M$^3$": **massively mixed methods**. We are, quite frankly, bringing several different approaches to bear in an exploratory way, looking for the most productive.

We've been applying quantitative methods to data from the student survey, the monthly reflection survey, and the daily implementation log. We're also quantifying aspects of teachers' TEFA practice, as captured in the classroom video, for quantitative analysis.

We've been applying qualitative methods to transcripts of the various interviews, open-ended responses to the monthly reflection survey, and select video segments. For this, we've been exploring theory-driven coding based on our prior research and experiences with TEFA. We've also been exploring a more grounded theory approach, applying open coding, microanalysis, and axial coding to the data.

Our collaborators at SRI International have recently begun discourse analysis of selected video segments.

Integrating all analytic tactics, our general strategy is to start by constructing detailed, rich **case study profiles** of each teacher's starting point and evolution. We want to understand their individual learning trajectories in as much detail as possible. As those profiles develop, we do **cross-case analysis** in an effort to identify general patterns and trends, distinct categories of experience, and central themes.

### research objectives

At this point, we've presented enough context that our official research objectives should make some sense.

Our first and primary goal is **to develop a better understanding of how teachers learn TEFA**, how they develop comfort with it, and how they adapt it to suit their needs. Although we're studying TEFA in particular, we believe TEFA is rich enough that much of what we learn will generalize to other pedagogies and instructional technologies.

Our second goal, building on the first, is **to improve our TEFA professional development**: to get better at helping teachers become happy and skilled with TEFA, faster and with less angst on their part.

Our third goal looks toward the future. We see a need for a rigorous, controlled implementation study of TEFA's measurable learning impacts on students and of the feasibility of large-scale TEFA professional development. To conduct such a study, we need **to develop knowledge and scalable instrumentation** that will allow efficient tracking of teachers' learning progress and implementation fidelity. Among other things, we need to know *when* in the course of a teacher's TEFA learning we should expect to see learning impacts.

# part 2: stories

Each teacher's experience of confronting TEFA and coming to terms with it is unique. It's a very personal journey, a narrative. It is tempting, but dangerous, to boil all our research down to a few concise bullet-points of generalized findings. We believe this would lose the essence of the teacher change process that we seek to illuminate. Therefore, we will tell four "stories" — narratives of the TEFA learning process for four of our participating teachers. Afterward, we will presenting some general findings, conclusions, and implications that cut across these cases as well as the others we are studying.

We've chosen this particular set of four four in order to illustrate four qualitatively different learning trajectories, all different and all illuminating. Every teacher's learning process is unique in at least some aspects, so it is difficult to call these four "representative". Nevertheless, they do illustrate a reasonable portion of the range of variability we have encountered.

**Tracy** is a very experienced, confident high school chemistry teacher. She took to TEFA like a duck to water. Even so, she did encounter tensions between TEFA and her teaching style.

**Kim** is an experienced high school math teacher, but new to this particular school. She struggled with TEFA and has had persistent doubts about its worth, but has kept working at it consistently, and has grown in the process.

**Renee** is a very experienced middle school science teacher whose personal style and outlook seemed to clash strongly with TEFA. Nevertheless, through cautious consideration and experimentation, she found creative ways to make it work for her.

**Gina** is a new middle school math teacher. She initially found TEFA to be thoroughly intimidating and got nowhere at all with it, but then she dramatically changed her approach to the project and took off — on her own terms.

### a. Tracy's story

"Tracy" is a high school chemistry teacher with over twenty years of experience. She is usually confident, positive and enthusiastic. At the outset of the project, her typical style of instruction was largely traditional and frontal. She expressed a strong need to be in control of classroom events, and in fact referred to herself as a "control freak." She struck us as an effective, competent traditional teacher. Student surveys suggest that she is very popular with her students. She has high expectations for her students, and wants them to be active and engaged learners.

Tracy started the project strongly enthusiastic about TEFA. Once the intervention began, she took to TEFA very naturally, and quickly demonstrated what we would judge to be high implementation fidelity. Nevertheless, she encountered several tensions between her perception of TEFA and her style, perspectives, and values. We can identify six specific tensions:

1. TEFA encourages teachers to help students' develop a deep understanding of concepts, but Tracy was more concerned with covering curriculum at a sufficient pace.

2. Good TEFA questions require time to craft, but — like most teachers — Tracy's prep time was limited.

3. TEFA seeks to foster student-driven dialogue, but Tracy's desire for control of the classroom led her to strongly mediate all classroom talk, with students interacting serially with her rather than with each other.

4. The typical TEFA question cycle respects the anonymity of students' answers to CRS questions, and relies on volunteerism in discussion participation, but Tracy's insistence that all students be engaged and participating caused her to frequently call on students by name to speak in discussions.

5. TEFA encourages teachers to be noncommittal in response to students' answers and let students struggle (up to a point) to figure things out for themselves, but Tracy was inclined to tell students the correct answer if she felt that they were taking too long to respond.

6. The final phase of the TEFA question cycle is for "closure," in which the teacher summarizes or otherwise helps students know what the point of the cycle was and what they should carry away from it, but Tracy had a self-identified weakness of not summarizing in general, and usually omitted this phase, leaving students unsure of what they were supposed to be learning.

Despite these tensions, Tracy's initial alignment with TEFA was good, and she struggled less with it than other teachers at that project site. She hit her stride quickly, and gradually made several adjustments to her practice that reduced these tensions.

One innovation Tracy introduced to TEFA was her practice of **scribing**: writing the key ideas from students' statements on her white-board, in order to keep track of the various ways of thinking raised and facilitate uptake of those ideas into later discussion. This practice had not been modeled in PD. Although it tends to keep students' attention focused on the teachers as the controller of dialogue, the practice seems to promote dialogicity in the discourse.

The initial tension between Tracy's desire to keep a rigorous pace in covering curriculum topics and TEFA's focus on **spending time** to thoroughly explore and solidify student understanding of central ideas was gradually mitigated as her views on the efficient use of classroom time changed. By the end of the first intervention year, she conceded that it was acceptable to budget more time for TEFA, claiming that developing students' reasoning skills was more important than covering content.

Similarly, Tracy's concern over taking **time to prepare** good TEFA questions subsided as she became more efficient at it, and also recognized how valuable the effort was. She was innovative in her question designs, developing styles more complex than those modeled in PD.

Tracy felt that one new type of question was effective in eliciting students' thinking and in her words, effective at "generating challenge to what kids understand."

Throughout the first year and into the second, Tracy was concerned about **student participation**: some students would not speak up in class discussion unless directly called upon. Later in the second year, however, she expressed satisfaction at their degree of voluntary participation and at the depth of their responses.

Tracy has wrestled with the challenge of **loosening control** of classroom activity and discussion, but she has tried different tactics for relinquishing at least some control without consequences that she considers unacceptable. As a result, we have seen a considerable increase in the frequency of student-to-student exchanges during whole-class discussion.

During the first year, Tracy made an effort to break her habit of quickly **telling students the answer**, and by the spring she admitted to us that she now liked not revealing answers, instead focusing on eliciting students' thinking. However, this change has resulted in a new tension between this practice and her students' desire to be told the answer.

Tracy's pre-existing, self-identified "weakness" of failing to summarize lessons for students was exacerbated by her initial extreme interpretation of our PD assertion that it is frequently useful to refuse to tell students the "right" answer to a question at all, instead letting the discussion of ideas speak for itself. In one videotaped class, she told her students that doing TEFA meant not being "allowed" to tell them the answer. Eventually, this misunderstanding was clarified, and Tracy made an effort to provide **explicit closure** at the end of each question cycle. Typically, she would tell students "how I would answer the question" to end the discussion.

Overall, as Tracy both adjusted her ways of implementing TEFA and evolved in her pedagogical views, most of the initial tensions she experienced were resolved and she became quite comfortable with the overall approach. During the second implementation year, she chose to focus her action research investigations on ways that she can use TEFA to motivate more students to do their homework and engage in non-TEFA aspects of the course. To us, this signifies that TEFA is no longer an "issue" for her or something that she wrestles with, but rather **a trusted tool** that she can use to address other challenges.

### b. Kim's story

"Kim" is a high school math teacher with seven years of experience before joining the project. However, she was new to this school and to the curriculum being used; the TEFA intervention began in her first semester with each. Not surprisingly, she was concerned about **being successful**, and being perceived as successful, in preparing her students for subsequent courses. She was interested in TEFA, but she had doubts and uncertainties about whether it would work in her class.

Our early installation of the classroom response system software at that school used a pre-release "beta" version, which had several bugs. This led to an early tension for Kim, as bugs and her own unfamiliarity with the software led to unexpected **technology glitches** that flustered her, thwarted her desire to conduct class smoothly and competently, and wasted precious class time. Within a month, however, the bugs were fixed, she had developed comfort with the software, and technology glitches were no longer an issue.

Another tension she wrestled with was a conflict between her **students' behavior** at this new school and her standards for how students should act. She found this new school to be far more relaxed about discipline than her previous one, and she was unhappy that when the bell rang to start a new period, students were still talking in the hallways rather than sitting in their seats. She disliked spending time settling students down at the beginning of every class.

This tension was not introduced by TEFA, but by her new situation. However, she figured out a way to use TEFA to address it: starting in November, she developed a pattern of beginning each class by posing a TEFA question, since she knew that students enjoyed these and wouldn't want to miss them. She discovered that she could use TEFA as a **classroom management tool**.

A third tension that Kim faced was between her doubt and uncertainty about **TEFA's worth** and her desire to be successful as a teacher. Since she wanted to try TEFA in her class but also wanted to be as successful as she had been in her previous school, she implemented the TEFA pedagogy in a safe, limited, conservative way. During the first semester, her questions tended to be simple and straightforward — what we call the "you know it or you don't" type. Unfortunately, such questions rarely lead to successful discussion.

Despite her caution, she had a **traumatic experience**. One student asked her to define the word "term" (of a mathematical expression), as used in a TEFA question, and she realized she didn't know how to define it. She had always assumed that the meaning was obvious to students. She struggled unsuccessfully to define it in class, and eventually gave up. In her journal, she described the experience as "panic, fear, frustration, anger, and embarrassment," and seems to have considered it a very negative event. From our perspective, however, it was a valuable learning experience for her, sensitizing her to the inherent ambiguities in even "clear" questions, the challenges of communicating reliably with students, the value of formative assessment to discover unexpected learning obstacles students face, and the challenge and importance of instructional agility.

As time went on and she had some rewarding experiences with TEFA, her **doubts eased**. Near the end of the first semester, she said that she "realized that trying it was fine; if it didn't work, ok, and if it did work I was ahead of where I had been." Also, she started to recognize the connection between the nature of the questions she posed and the quality of the resulting discussion.

At the start of the second semester, Kim had **an eye-opening experience** while teaching with a question that she had developed in PD. She wrote in her journal that "The result was amazing. Students boldly shared opinions and gave sound reasoning and examples to support their ideas… For the first time ever I experienced with my students what I would consider to be the 'beauty' of using TEFA/PRS. It really worked!" This proved conclusively to her that TEFA could generate strong student discussion and change the class atmosphere, and that she, personally, could be successful with TEFA. It also cemented her belief that the key to generating good discussion was developing good questions.

During this second semester, we see evidence that her overall view of teaching evolved, becoming more "**student-centered**." In our baseline interview, she said that her responsibility as a math teacher was "to convey information in such a way that it is easy to learn by as many students as possible." But one year later, she said, "I realize that I was planning it from the standpoint of what I need instead of what the students need." She also changed the way she approached lesson planning. She said, "TEFA makes me think about the big picture, especially when I'm introducing a topic. It makes me come up with a stimulating question… that provides that connection…"

Although some of her tensions with TEFA had been resolved, she still struggled with others. For one thing, she was now very concerned with creating "effective" questions that would enhance class discussion, but found that this required more **prep time** than she could easily afford to spend. For another, she found that good classroom discussion took considerable class time, and was concerned about how that might affect her ability to **cover content** at a sufficient pace.

These concerns continued into the second intervention year as Kim persisted with TEFA. Another concern that grew in her mind was dissatisfaction with poor **student participation** in whole-class discussion, and concern whether TEFA was of much benefit to them. In particular, she noticed that boys were more likely than girls to speak up. So, she created and administered a survey for her students, inquiring about their perceptions of how valuable different kinds of learning activities — including whole-class discussion, small-group discussion, homework, reading, and the like — were. She reported that "their perception is that whole class discussion is not as helpful a way to learn as small group work…" Interestingly, she found that boys were more likely than girls to value whole-class discussion, whereas girls preferred small-group problem-solving.

As a result, she adjusted her use of TEFA, placing more emphasis on small group discussion and using whole-class discussion primarily to have students "report out" the results of their small-group work.

As of the second semester of the second intervention year, Kim is still struggling with the pressure TEFA exerts on both her preparation time and class time. She tells us quite frankly that using TEFA is valuable, but she's not sure it's worth the time required to create good questions week after week. Nevertheless, we see that her perspectives and practices have changed significantly over the past year and a half. She has become more student-centered, focusing on students' thinking and needs and modifying her lessons and teaching practice based on what she learns from her students' responses. In other words, formative assessment has become **an integral part of her teaching**, and seems likely to remain so whether she uses TEFA itself or not.

### c. Renee's story

"Renee" is a middle-school science teacher who, like Tracy, has over twenty years of experience. She displays confidence in her teaching abilities. She is respectful of her students, and **firm and clear** about her expectations. Her interactions with students are concise and highly structured. She expresses and manifests a strong concern with shaping students' social skills.

At the outset of the project, Renee seemed curious and optimistic about TEFA. However, she seemed to find our initial professional development workshop unpleasant, exhibiting **severe reluctance** to either be videotaped (we taped the entire workshop, for our own records), or to practice-teach a mock TEFA lesson with her peers in the role of students. As far as we can tell, she simply hates to be "on the stage" or the object of attention. This is consistent with her teaching style, which relies on student seat-work and almost never involves frontal instruction.

Renee acquiesced to our normal videotaping schedule of four classes during the year, but she did not use TEFA during a taping for the first two visits, despite our attempts to schedule taping to coincide with TEFA. Her logs and journal indicate that she was using TEFA when we weren't present. In fact, she documented her TEFA usage in meticulous detail, including **highly reflective comments** on both the successes and the failures. However, she rarely spoke during PD meetings.

Renee also struggled with a tension between what she believed her students capable of, and what she thought TEFA expected of them. She considered her students to be "concrete thinkers" unable to handle the abstract reasoning and argumentation she had seen modeled in PD meetings. She noted early that her students' attention span was short and their focus poor, so that they had trouble staying engaged in lengthy TEFA discussions.

Her solution was to **restrict TEFA usage** to one or two days a week, for a maximum of fifteen minutes at a time, and to keep her questions relatively simple. This introduced a new tension between what she was doing and what she thought PD staff wanted and would consider "acceptable," but she was willing to live with that.

Another tension Renee encountered was between the format of our PD program and her needs. In her monthly surveys, she regularly complained that she expected **more time** to create TEFA questions and discuss them with peers. This tension was never resolved to her satisfaction, partially because we were backlogged with data and did not see these complaints in time to address them.

This state of affairs lasted for several months, through the fall semester of the first intervention year and into the spring. During that period, she **gradually expanded her repertoire** of uses for TEFA. Each month she would introduce one or two new styles of question or ways of using TEFA, often inspired by examples and ideas from the PD program. She generally reported success and satisfaction with the outcome, but remained cautious and reserved in her embrace of TEFA. During the third quarter of the year, she finally allowed us to videotape her doing a question cycle with the CRS.

During the spring, Renee was concerned about her students' decreasing participation in TEFA discussions. About half-way through the term, she engaged her classes in a frank, meta-level conversation about the problem. They told her they'd prefer not to see the histogram of responses until after they'd discussed a question. In response, she **modified her implementation** of the question cycle: she'd pose a question and collect answers, but not show the histogram. Instead, she'd let students discuss the question as a whole class. Then, she'd collect a second round of answers, present both histograms, and let students discuss how and why their answers had changed between the rounds of polling. This pattern seemed to be more comfortable for students, and Renee reported that participation levels rose dramatically.

Shortly thereafter, Renee tried an innovative solution to the tension she had experienced between her dislike of frontal instructional modes and the canonical TEFA pattern of having the teacher moderate whole-class discussion. She formulated an explicit set of rules for whole-class discussion, and presented them to her students. Then, during the whole-class discussion phase of the TEFA question cycle, she would physically and verbally **remove herself from the group** and allow the students to self-moderate, responding entirely to each other and setting their own direction. The experiment was successful enough on the first try that she allowed us to record the second on our fourth and final videotaping visit of the year.

Unfortunately, a new **technical problem** prevented her from using her CRS for the last month of the year. Without warning, her classroom computer was replaced with a new and more powerful one, but without the CRS software installed. Because our PD and data collection schedules for the term were over by this time, we did not find out about the problem until much later. Her final monthly survey expressed irritation that she wanted to use her CRS and could not.

Overall, Renee manifested a **remarkable transformation** over the first year. She began as a teacher doubtful about TEFA's suitability for her students, very conservative in her use of it, and deeply uncomfortable with core portions of the pedagogy. By the end of the year, she had developed a wide range of TEFA uses and question styles, found ways to elicit quality student participation, and reconciled her enactment of TEFA with her values and views. She came to value the formative assessment information TEFA provides, and added a new mode of classroom interaction — the self-moderating whole-class student discussion — to her repertoire.

Unfortunately, two major tensions remained unresolved. One was the conflict between her deep dislike of being scrutinized and the project's need to conduct research on her learning process. The other was her continuing feeling that PD was not meeting her needs. She found a way to rectify those, too: before the second intervention year began, she sent us a terse email stating that she was exercising her right to **drop out** of the project, that she would like to keep her CRS if we would allow, and that she did not want us to contact her for the rest of the summer. She has not responded to email from us since then.

So, as far as we know, Renee may still be practicing some form of TEFA, but she has dropped completely off our radar.

### d. Gina's story

Gina teaches middle-school mathematics. When we began our intervention, she was starting **her second year** as a teacher, and her first year with a new curriculum. Her incoming class of eighth graders was rumored to be unusually difficult. As if all that wasn't challenge enough, she was dealing with some complex and emotionally draining personal issues that would demand a lot of her attention and energy for the next year or more.

At the beginning of the school year, after the summer workshop, Gina was generally optimistic about TEFA. She was a little concerned about how well it would work with her particular students, and about her ability to operate the technology. Nevertheless, she had **high hopes** for diving head-first into TEFA and mastering it.

Like most participants, she quickly got the hang of using the technology's essential features, and initial tech glitches were resolved within a month. However, she discovered very quickly that TEFA is far **harder to learn** and execute than she'd imagined.

For the rest of the fall semester, she struggled with two general tensions between her picture of TEFA and her assessment of the reality of her classroom. One had to do with how well TEFA "fit" her students. As she saw it, TEFA expects students to engage each other and the teacher in rich, thoughtful, extended discussion and argument about abstract ideas. However, she saw her students as "concrete thinkers" with limited ability to reason abstractly, poor social skills, and little willingness to engage thoughtfully in discussion. This led her to **struggle to find questions** that were accessible to her students, and yet in the spirit of TEFA. She spent many hours vainly searching the web for question ideas. It also led her to struggle with managing behavior and eliciting participation.

These struggles were exacerbated by a second tension, having to do with expectations. She formed an extreme view of what successful TEFA should look like. For example, PD staff had argued that ambiguous questions, and questions with more than one defensible answer, are pedagogically efficacious. She interpreted this to mean that TEFA questions should *never* have one unambiguously right answer. Similarly, she believed that *all* questions must be deeply conceptual, and *all* questions must provoke rich and extensive discussion among students. It seems that in our attempts to paint a vision of TEFA that contrasted with traditional instruction and showed what was possible and worth aspiring to, we had accidentally created a **forbiddingly high bar** for Gina.

Thus, Gina struggled greatly trying to find or create questions that would meet this self-imposed standard. Week after week, we would present new question styles in PD and assign "homework" asking teachers to create their own questions of that type and try them out in class. And week after week, Gina would find herself unable to do that in a way that she could see working with her students. Not surprisingly, she became quite **frustrated**, and her self-confidence as a new teacher took a severe hit.

Over the December holidays and January, no project PD occurred for nearly two months. During this time, Gina recognized that her own fears about using PRS "well enough" were a barrier to her. She realized that if TEFA was going to work for her, she would have to do it for herself and her students, not for us. For the rest of the spring semester, Gina adopted a **new attitude** and resolutely ignored all professional development requests to do or try specific things with TEFA, even though she feared this might harm her grade when she got graduate course credit for the program. Instead, she explored different ways of using TEFA, guided by her own ideas and discussion with peers instead of professional development. And she started thinking of PRS more as a tool she could use for her own ends, and less as a "way" that she had to follow.

One of the things she found was a "niche" for TEFA: a way of using it that seemed manageable, reliable, and productive for her, even though it was a pale shadow of the TEFA she'd seen presented and modeled by project staff. For her, TEFA became a way to have students **review questions** for the upcoming state-wide math exam, and to find out more about what topics on that exam they needed help with. She largely ignored the "dialogical discourse" and "question-driven instruction" aspects of TEFA, instead focusing on a relatively patterned, but useful, formative assessment role.

She also experimented, cautiously and modestly, using TEFA in different roles. She had some success experiences that motivated her to keep going. Over the spring semester, her **confidence rebounded**, and the degree of stress she felt regarding TEFA was drastically reduced. She still perceived a tension between TEFA as it was portrayed in PD meetings and TEFA as she was practicing it, but she was willing to live with that.

Another tension that persisted was between her students' ability to behave responsibly in **small-group work** and Gina's desire to have more small-group experiences for them, both with and without TEFA. She persevered and had some limited successes with this. She also noticed that some classes were much better at small-group work than others, which reassured her that the source of the problem was with the students, not with her competence.

After the summer break, Gina confidently fell into a pattern of using TEFA every day to review for the state-wide exam. For the entire fall semester, however an intermittent and unpredictable **technical problem** with the school network prevented her from using her computer, and thus her CRS, perhaps one day a week or more. She found this very distressing. Interestingly, one year ago she had struggled to make herself use TEFA at all; now she felt hamstrung if she couldn't start every class with it.

Gina also continued **experimenting** with TEFA in different roles, gradually expanding her repertoire of ways she felt comfortable using it. She started spontaneously inserting "on-the-fly" questions, in order to gauge student understanding of something that had just occurred. She started adjusting instruction in real-time based on student responses. She linked TEFA to hands-on activities, using pairs of questions around an activity to first motivate it and then later assess and solidify learning from it. She made status-check questions a regular part of her TEFA pattern, asking students how confident they felt about a topic. She even began sprinkling in questions designed to foment discussion or provoke wrestling with challenging ideas — exactly the use of TEFA that had so intimidated her a year ago.

In other words, she couldn't handle TEFA as the whole big intimidating entity she initially perceived it as. But by rejecting everything except a narrow and limited aspect, developing confidence and skill with that, and then gradually extending **at her own pace**, she has ended up making a relatively rich and varied implementation of TEFA an essential part of her instruction. And in the process, she seems to have raised her perceptions about what students are capable of handling, sharpened her appreciation of formative assessment, developed her ability to make agile adjustments in class, and become much more at home as a teacher.

# part 3: findings

### a. some observations

It is clear to us from the preceding four stories and from the remainder of our data that TEFA is hard, **change is hard**, TEFA can be intimidating, and that teachers need to develop a range of new skills to succeed with it. It is also clear that **teachers *can* change** and learn TEFA, and they can make significant progress in this over course of one year of professional development. In the lessons that we videotaped and analyzed from our first school, we have seen several changes exhibited by most teachers:

• the **time spent in discussion** during the question cycle increased;

• the **frequency of student-student interactions** during discussions increased;

- teachers' use of **IRE types of interactions** decreased, especially in the earlier parts of question cycles;

- teachers increased the variety of **discussion formats** that they used as part of the question cycle.

Furthermore, several teachers reported that the use of a CRS and TEFA helped them **gain more information** about how their students were thinking about concepts. And they reported (verified by videotape data and pre/post-taping interviews) instances where they **changed their lessons** as a result of the formative information that they gained about students' thinking.

The stories above depicted each teacher's coming to terms with TEFA as a path or "trajectory" he or she follows. One aspect of this trajectory is the skills learned and knowledge gained at each point in time. Individual teachers' learning trajectories are personal and idiosyncratic, but trajectories seem to share certain general features.

Earlier, we indicated that the skills necessary to master TEFA can be grouped into **five general skill areas**: using the technology, creating questions, orchestrating classroom discourse, understanding and adjusting to students, and integrating TEFA with all other constraints and context. It seems that teachers typically encounter and wrestle with the first three skill areas in that order. Operating the technology and dealing with glitches looms large to them at first, but most become comfortable with that within a month. Then, they realize that designing effective questions is more challenging than it at first appears, and spend much of the first year (or more) focused on this. Slightly later — usually before question design has been "mastered," but after some degree of comfort has been reached — teachers become concerned with the dynamics of their classroom discussion, and reconsider their role and practices in that.

The fourth skill area, which we colloquially refer to as "getting inside students' heads, and then knowing what to do about it," seems to stay implicit for the first year or more of TEFA learning. It is a factor, but not one that most teachers explicitly focus on. However, we have known at least one extremely proficient, experienced TEFA practitioner who claims that if he focuses on understanding what his students are thinking, everything else falls into place.

The fifth skill area, integrating TEFA with "everything else," is more idiosyncratic. Teachers' contexts (and their perceptions of those contexts) vary drastically, so both the degree and the nature of the challenge this skill area poses to them varies equally drastically. And, because context tends to change, this area must frequently be revisited. In general, this seems to be a skill area teachers worry about intermittently, and more strongly at the beginning of each academic term.

Our four stories ended with the teachers in **very different places** on their trajectories. Tracy has become an expert user of TEFA, and we believe that she has incorporated it deeply into her way of being a teacher. Kim has steadily increased her proficiency in TEFA, and her conception of TEFA pedagogy has become increasingly similar to ours, but it is still a struggle and she has doubts about whether it is worth the time it requires. Gina has transformed TEFA into pedagogy that she is comfortable with and is using it extensively in her classes, in an increasing variety of roles. Finally, we saw that Renee has developed an understanding of TEFA consonant with ours, has had success using it in her classes in a careful and limited way, but has disassociated herself from the research project. One of the dominant findings of our study is that "all teachers are individuals," and the TEFA learning trajectory is idiosyncratic.

The TEFA learning trajectory is also **bumpy**. What we mean by this is that the changes in the teachers and in their pedagogy does not happen smoothly, as a continual and gradual progression. Rather, their conception of TEFA, their ways of using it, and their facility with it can vary considerably and shift suddenly as they wrestle with tensions and dissonances, experience seminal events, have insights, and get input from their peers and from the PD program.

Another phenomenon we have observed is that teachers need to find an *entry point* for TEFA: a way of using it and a niche for it within their practice that is within their "zone of proximal development" (ZPD, Vygotsky, 1978). That is, they must be able to envision a way of using TEFA that they believe is within their ability, is within their students' capacities, suits their subject matter and level, integrates with their overall teaching patterns, and will produce worthwhile results. Tracy found an easy entry point using TEFA for whole-class discussions about key chemistry concepts; Kim eventually developed one using TEFA to introduce new topics at the start of a class period; Gina struggled, feeling that TEFA was completely beyond her ZPD, but ultimately discovered one using TEFA for standardized exam preparation; and Renee had various success experiences, but did not seem to find a comfortable entry point or niche for TEFA within the year we studied her.

We have found that helping teachers learn TEFA gets more challenging as teachers get farther from the **PD staff's own background**, which is university-level physics instruction. We have more trouble communicating productively with middle school teachers than with high school teachers, and more trouble with mathematics or biological science teachers than with physical science teachers. The result is that many of these groups feel more "left on their own" to figure out TEFA and create effective questions. Differences in language are one source of difficulty: physicists talk about "problem solving" where biologists might talk about "model-based reasoning," and math teachers talk about "problem solving" but typically mean something more procedural and less conceptual. Differences between our pedagogical content knowledge (PCK, van Driel, Verloop & de Vos, 1998) and theirs is another source of difficulty: we are better at generating compelling examples of CRS questions or responses to student questions in subjects we have taught ourselves.

### b. co-evolution model

Looking more closely at stories of teacher change from the project, we have realized that a useful way to see and describe the process is in terms of:

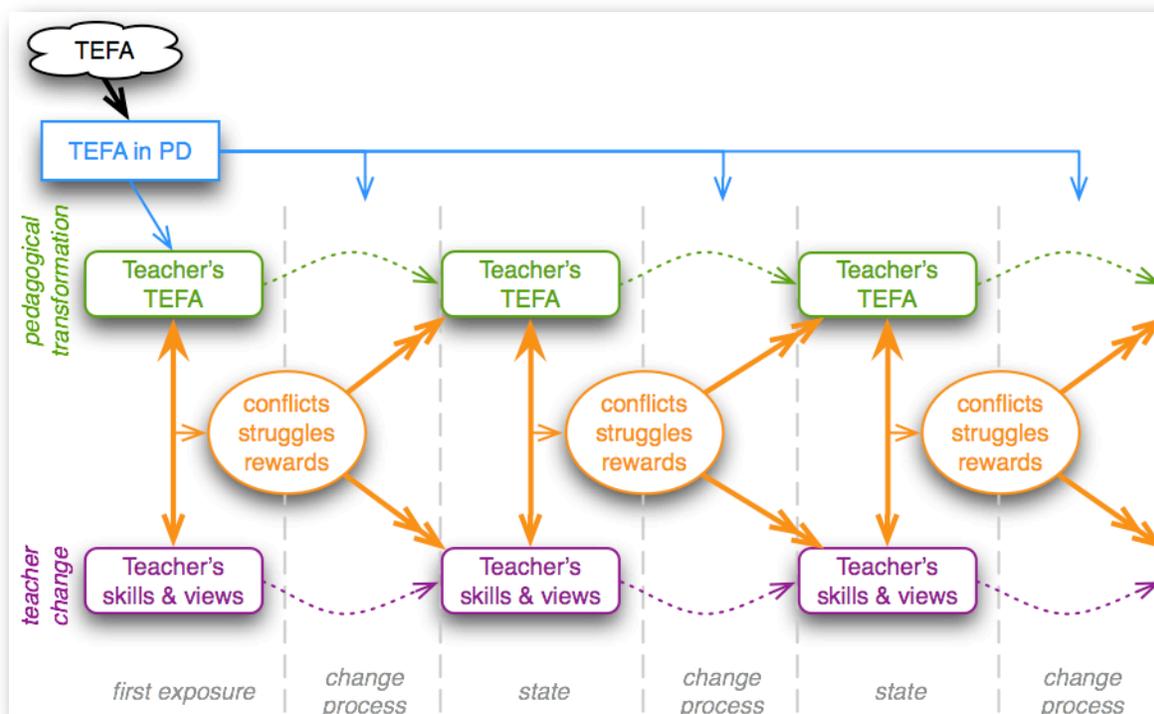

*Figure 3: Graphical representation of a model for the co-evolution of teacher and pedagogy.*

- the **alignment or misalignment** (perceived as tensions) between a teacher's skills, views, and context, on the one hand, and her conceptualization of TEFA and attempts to enact it on the other;

- the **conflicts, struggles, and rewards** she experiences as a result of these alignments and misalignments;

- the **changes to her conceptualization of TEFA** and to her ways of attempting TEFA that occur in response to these conflicts, struggles, and rewards; and

- the **changes to her skills, perspectives, and general "way of being a teacher"** that also occur in response to the conflicts, struggles, and rewards.

This has led us to formulate our "model for the co-evolution of teacher and pedagogy" (figure 3). This model is the result of some very recent thinking on our part, and should be considered a work in progress.

TEFA, represented at the far top left of the figure, is the pedagogy that we hope the teachers will become skilled at and internalize, recognizing that exactly how they practice it will vary depending on who they are and how they perceive their context.

TEFA as represented in PD is not identical to TEFA itself; consciously and unconsciously, we stress some aspects, omit others, and communicate it in specific ways in response to the details of the PD situation and our assessments of the teachers and their needs. This representation of TEFA evolves over time as the PD program continues and we "adjust the message" in response to our observations of the teachers.

The understanding each teacher forms of TEFA is not identical to the TEFA we attempt to present, either, since all they experience will be filtered through the lens of their own preconceptions, sensitivities, and needs. Furthermore, the TEFA they aim to implement will be yet different, as they choose some portions, discard others, and generally adapt this **interpretation of a representation of TEFA** to fit their needs.

What the teacher brings to this — to the version of TEFA they aspire to implement — is everything that makes them who they are as a teacher: their values, beliefs, perceptions, and skills. In every case we have seen, this "who they are as a teacher" has some points of misalignment, some **tensions**, with TEFA as they understand it and aspire to practice it. Tensions result in **conflicts and struggles** that the teacher experiences in their practice of TEFA. We see these conflicts and struggles as the "**driving engine**" of teacher change: in attempting to resolve them, the teacher is motivated to either evolve their understanding of TEFA, adjust their ways of implementing TEFA, or grow in their views and skills.

Points of good alignment between teacher and TEFA — the opposite of "tensions" — can produce **rewarding experiences**, often unanticipated. These can also drive change as the teacher seeks to capture or maximize them, and they can confirm the teacher and provide motivation to continue with the struggle.

Thus, the model represents teacher change as **a series of transitions** or "change process" steps between states. The transitions are depicted as discrete and sequential, but they need not be; we have seen both gradual evolution (as in Tracy's case) and sudden change (as in Gina's). Ongoing professional development serves to help **mediate** these transitions, both by provoking additional tensions (for example, by "raising the bar" in terms of what quality TEFA implementation ought to aspire to), and by suggesting resolutions (for example, by teaching specific skills or suggesting alternative perspectives).

To us, a dominant and very significant feature of the model is that it can be viewed as **an evolving dialectic** between three "narratives." One, which is indicated only very generally and vaguely in figure 3, is our narrative as professional development staff: the evolution in our understanding of our participating teachers, the development of our PD skills and tools, and the "story" we present in PD. We intend to examine this narrative further and building a more detailed model of it in subsequent project work.

The other two narratives are internal to the teacher. One, which we have labeled "**teacher change**," is the story of the teacher's growth as a practitioner: acquisition of new skills, development of new perspectives, realization of new or newly emphasized values, and the like.

This is the narrative we had been focused on when we began the project; we thought of professional development as helping teachers to develop skills and change views.

The second internal narrative, which we have labeled "**pedagogical transformation**," is new to our thinking. Although we acknowledged that every teacher would implement TEFA slightly differently, we desired to minimize the differences and aspired to good "implementation fidelity." Quite frankly, it was rather horrifying for us, TEFA's developers, to watch teachers such as Gina and Renee take a machete to TEFA, hacking off parts we saw as essential in order to come to terms with it — or, rather, in order make TEFA come to terms with them. But we have come to see such customization and personalization of TEFA as essential to the teacher learning process.

Thus, one can view the model vertically (in the figure), as a progression of states and transitions, where states are defined by tensions and their consequent conflicts and struggles, and transitions are defined by changes to the teacher and her enactment of TEFA. Or, one can view it horizontally, as two ongoing narratives, one describing change in the teacher and one describing change in her understanding and enactment of the pedagogy; the narratives are connected dialectically by the tensions and dissonances between them at any point in time.

### c. implications

Appreciating the significance of the "pedagogical transformation" narrative to the teacher change process has produced **a shift in our thinking** that goes much deeper than simply admitting that teachers will adapt our pedagogy, and accepting it as a necessary evil of professional development. We realized that we had implicitly been trying to recast our participating teachers in our own image, and make them our proxies in the classroom, teaching the way we would. Recognizing the necessity of the pedagogical transformation narrative in our model of the teacher change process, we are coming to think of our goal more as to help each teacher blossom in his or her own way, by providing a rich and powerful pedagogy *that they can and should adapt to their own strengths, needs, and outlook*. Language along these lines was always in our talk, but now we realize its implications much more deeply.

(In retrospect, this should have been obvious. Even those of us closely involved in developing, refining, and articulating TEFA do not implement it identically.)

This new perspective also challenges the utility of the "implementation fidelity" concept. Against what should a teacher's TEFA implementation be compared, if the goal is not to have all teachers approach some hypothetical ideal implementation? A canonical reference implementation may be necessary for a study that seeks to establish the student learning impacts of TEFA, but measurements of fidelity must be suitably lenient in non-essentials.

We end with some other **implications for in-service teacher professional development** that arise from our co-evolution model or from our general findings and experiences to date.

- For PD to succeed, it must be **sustained over time** as teachers follow an extended process of personal change and pedagogical transformation.

- PD should be structured so that teachers have **multiple entry points** into the new practices, so that what they are exhorted to try does not exceed their ZPD.

- Given the impact of critical events in teachers' adoption of new practices, PD should help teachers **recognize seminal experiences** for what they are, and to help them use evidence about those events as a way to see that their new pedagogy is "sensible, beneficial, and enlightening" (Feldman, 2000).

- While PD should be sustained, **breaks from PD** (e.g., school holidays or summer vacation) can help teachers reassess their practice and reflect on the implications of what they have been learning.

- PD staff should monitor the **tensions and dissonances** that the teachers experience and provide strategies for eliminating or mitigating them, so that resulting conflicts and struggles do not seem insurmountable. Similarly, staff can help teachers **recognize the rewards** of new practices.

- Developing high-quality curriculum (CRS questions) for TEFA has been a major difficulty and source of stress for teachers. If possible, PD should provide at least a minimal core of curriculum materials suitable for each teacher's subject and level, so teachers can become familiar with the approach without the burden of curriculum design. At the very least, a wide **range of exemplary materials** are necessary, varied in subject, level, style, and goals.

- A PD program must strike a careful balance between **careful design** to address all requisite skill areas in an effective learning sequence, and **attentive responsiveness** to teachers' idiosyncratic and changing needs.

- Finally, the staff of a PD program should recognize the difficulty of communicating across disciplinary and teaching-level boundaries, and seek collaborators or other resources to help bridge gaps.

### d. the future

At present, we are slightly more than half-way through the three-year intervention at our first site, still in the intensive first intervention year at our second site, and still collecting baseline data at our third and fourth sites. This means all findings herein should be considered preliminary.

In the future, we will be collecting data to substantiate these findings. We will also be modifying our PD design for sites 3 and 4, in order to take advantage of and test our co-evolution model of teacher change. With additional case study analysis, we hope to develop a more comprehensive description of teachers' TEFA learning pathways and stages, perhaps identifying distinct categories of teacher, common entry points, typical stumbling blocks, and the like. Further on in the project, we will turn to the development of more scalable PD and more scalable research instrumentation, as we look towards a controlled study of student learning impacts due to TEFA.